\documentclass[conference]{IEEEtran}

\pdfmapfile{+txfonts.map}

\usepackage{amsmath}
\usepackage{txfonts}

\usepackage{color}
\usepackage{graphicx}
\usepackage{subcaption}
\usepackage{framed}
\usepackage{url}

\usepackage{flushend}

\begin{document}

\newcommand{\Fix}[1]{\textbf{[[}{\color{red}#1}\textbf{]]}}

\newcommand{\eg}{e.\,g.}
\newcommand{\ie}{i.\,e.}
\newcommand{\etal}{et al.}

\renewcommand{\implies}{\,\Rightarrow\,}
\newcommand{\ReqSpec}{{\cal R}}
\newcommand{\Reqs}{\mathsf{Reqs}}
\newcommand{\req}{\mathsf{req}}
\newcommand{\parent}{\mathsf{parent}}
\newcommand{\Types}{\mathsf{Types}}
\newcommand{\type}{\mathsf{type}}
\newcommand{\link}{\mathsf{link}}
\newcommand{\TestSuite}{{\cal T}}
\newcommand{\tc}{\mathsf{test}}
\renewcommand{\ell}{\mathsf{xpctd}}
\newcommand{\TCs}{\mathsf{Tests}}
\newcommand{\As}{{\cal E}}
\newcommand{\Platform}{{\cal P}}
\newcommand{\Expf}{\mathsf{Xpctd}}

\newcommand{\TestStatus}{{\cal S}}
\newcommand{\Success}{\mathsf{Success}}
\newcommand{\Fail}{\mathsf{Fail}}
\newcommand{\Undefined}{\mathsf{Undefined}}

\newcommand{\Satisfied}{\mathsf{Satisfied}}
\newcommand{\Violated}{\mathsf{Violated}}
\newcommand{\Unknown}{\mathsf{Unknown}}
\newcommand{\NoResult}{\mathsf{No~result}}

\newtheorem{definition}{Definition}

\title{A Logical Approach to Generating Test Plans
}

\author{
\IEEEauthorblockN{Tobias Morciniec}
\IEEEauthorblockA{Daimler AG, Stuttgart, Germany\\
tobias.morciniec@daimler.com}
\and
\IEEEauthorblockN{Andreas Podelski}
\IEEEauthorblockA{Albert-Ludwigs-Universit\"at Freiburg, Germany\\
podelski@informatik.uni-freiburg.de}
}

\maketitle

\begin{abstract}
During the execution of a test plan, a test manager may decide to drop
a test case if its result can be inferred from  already executed
test cases.
We show that it is possible to automatically generate a test
plan to exploit the potential to justifiably drop a test case and
thus reduce the number of test cases.
Our approach  uses Boolean
formulas to model the mutual dependencies between test results.
The algorithm to generate a test plan comes with the formal guarantee of
optimality with regards to the inference of the result of a test case from
already executed test cases.
\end{abstract}
\bigskip


\section{Introduction}
\label{sec:introduction}
In this paper, we present the theoretical investigation of a
question which is potentially of practical relevance.   
The question is whether
automatically generated test plans can help with the problem
that, in practice, we always need to reduce the number of test cases
(see, e.g.,~\cite{Andres12,Gotlieb14,Saha15}).
The problem is exacerbated
in system testing in the automotive industry (see, e.g.,~\cite{Broy07}).
A single test case for a high-level test platform such as HiL
(Hardware-in-the-Loop), SiL (Software-in-the-Loop), or the vehicle
itself, typically involves complex test setups with hours of human
labor and hours of execution.
Every single opportunity to justifiably drop a test case is valuable.

Recent work~\cite{ArltMo15} shows that one can
exploit logical
dependencies between system requirements in this context.  The idea here
is that, during the execution of a test plan, a test manager may
decide to drop a test case if its result can be inferred
from the already executed test cases.  The
inference of this logical \emph{redundancy} of a test case can be done automatically.

The redundancy of a test case depends on the ordering of the test
cases; it may be inferred in one ordering, but not in another. 
To give a simple example, we consider the logical dependency between
the requirement $\req_0$: \emph{rain sensor detects rain} and the
requirement  $\req_1$:  \emph{sunroof
closes automatically when it is raining}.  Given that the rain sensor
provides the only way to detect rain, the second requirement can only
be satisfied if the first one is,  formally
$\req_1\Rightarrow\req_0$.  We assume that the requirements are linked
with the test cases $\tc_0$ and $\tc_1$, respectively.  Assume that
both test cases fail.  If the test plan fixes the order
$[\tc_0, \tc_1]$ then $\tc_1$ is inferred to be redundant after the execution of
$\tc_0$.
 If, however, the test plan fixes the order
$[\tc_1, \tc_0]$ then no test case becomes redundant.  

The reader may have spotted an immediate issue here.  In the case where the test cases
$\tc_0$ and  $\tc_1$ succeed, the test plan with the inverse ordering will be
optimal (i.e.,
 if the test plan fixes the order $[\tc_1, \tc_0]$
then $\tc_0$ becomes redundant after the execution of
$\tc_1$, and if the test plan fixes the order
$[\tc_0, \tc_1]$  then no test case becomes redundant). 
Apparently, in order to know what test plan provides an optimal
ordering, we first have to execute the test cases.

The solution to the issue lies in the fact that a
test manager, when trying to find an optimal test plan, has in mind
her (more or less vague) \emph{expectation} of what the results of the test
cases will be. We will optimize the test plan with respect to the
expectation of the test manager.  In other words, the test manager
will no longer need to find an optimal test plan by herself; instead,
she specifies her expectation and the optimal test plan will be generated
automatically from there. 

To specify her expectation, the test manager can
start with a \emph{default specification} (see Section~\ref{sec:UE}).
She can simply take one of the three specifications as is
or she can take it as a basis and change it for individual test cases 
according to her personal insight and experience.  
She may also leave unspecified her expectation on a test case.
An unspecified expectation (or one that turns out to be wrong) may
possibly lead to missing a redundancy (in the case where the redundancy could have
been inferred otherwise) but our
formal criterion for the optimality of a test plan will still apply.

In the example,  if the expectation is that both test cases
fail, then the test plan with the optimal ordering is
$[\tc_0, \tc_1]$.  If the expectation is that both test cases succeed,
then the test plan with inverse ordering will be optimal.   (The
expectation that one of the two fails and the other one succeeds would
not be compatible with the logical dependencies between the
requirements linked to the test cases; in that case, we would ignore
the expectation for the two test cases for the purpose of generating a
test plan.)

In our example, it is easy to generate an optimal test plan from the user's
expectation about the test results. 
In general, however, this cannot be done manually.
The dependencies between requirements are  complex (between more
than two requirements), each test is linked to more than just one
requirement, there are dependencies due to the different test
platforms, and finally we have to take into account the expectation of
the test manager.  It is \emph{a priori} not clear whether it is always
possible, given an expectation on test results, to order the test
cases such that the number of redundant test cases (and thus the
number of opportunities to drop a test case) is optimal when the execution of
the test cases yields the expected test results.  Even if this is the case, it seems impossible to
manually find a test plan with an optimal ordering, but it is \emph{a
  priori} not clear that one can automatically generate such a test
plan.

Our approach is to encode the expectation of the test manager in logical
formulas in such a way that we can derive conditions on an optimal ordering by logical
reasoning.  The formulas are Boolean; the logical reasoning can be automated by
calling a standard SAT solver.  We add these logical formulas to the logical
formulas that we use to formally model dependencies between results, links
between test cases and requirements, and dependencies due to the different test
platforms.  We can prove a fundamental property of the resulting algorithm: its
completeness.  If there exists an ordering of test cases in which the result of
a test case becomes redundant (in an execution of the test cases with the
expected results), then the algorithm will infer such an ordering.

Generating a test plan is a difficult task which depends on many,
often contradictory goals.  Solving the task will always eventually
rely on the human, her experience, insight, and intuition.  Automatic
support aims at helping the human to concentrate on the conceptual
complexity of the task, by removing at least some of the burden of the
combinatorial complexity.  Our work is a first step in this
direction.  Using the logical approach to generate test
plans allows the user to concentrate on specifying her expectation on test
results, i.e., the one parameter that defines the specific
optimality criterion, and frees her from the complexity of the
above-mentioned dependencies.

\section{Preliminaries}
\label{sec:preliminaries}

In Section~\ref{sec:GTP}, we will explain how one can automatically generate
test plans that are optimized for the possibility of identifying
redundant test cases.
In this section, we will explain how one can automatically identify when a test
case is redundant (so that it can potentially be dropped without executing it).
Thus, the purpose of this section is to make the paper self-contained. The
technical content applies generally to every setting where the redundancy of a
test case can be inferred from already executed test cases.
The material in this section covers mostly work from~\cite{ArltMo15}. This work shows how
the documentation of requirements with links to tests on different test
platforms can be exploited for identifying redundant test cases.

What the reader should take away from this section is the following.  We
introduce a Boolean variable $\tc$ for each test case; the Boolean value (true
or false) for $\tc$ stands for the result of executing the test case (the test
case succeeds respectively fails). The result of a test case can become
redundant, from the present test status $\TestStatus$, i.e., the results of a
set of other test cases (which have already been executed), and from the
specific setting which is given by: the requirements specification $\ReqSpec$,
the test suite $\TestSuite$, and the test platforms $\Platform$. The redundancy
can be mechanically inferred (using a SAT solver), namely by logical inference
of $\tc=true$ resp.\ $\tc=false$ from a logical formula $\ReqSpec \wedge
\TestSuite \wedge \Platform \wedge \TestStatus$.
The logical formula is the conjunction of logical formulas which model the
requirements specification, the test suite, the test platforms, and the present
test status.
\smallskip

\subsection{Logical Dependencies between Requirements}
\label{subsec:dependencies_between_requirements}
Our approach is generally applicable in every context where one can
derive logical dependencies between requirements.  For example, it is
possible to use a formal
specification of an ontology as in~\cite{Wang07}.
For concreteness, we will describe the setting of~\cite{ArltMo15}.
Here, the dependencies are derived (mechanically)
from the documentation of (informal) requirements in a structured
format  (a format which is amenable to mechanical processing;
see~\cite{FilipovikjNR14} and~\cite{YeganefardB11}). 

The requirements document organizes
the requirements in a \emph{hierarchy} and classifies requirements on
the same level in the hierarchy further by
assigning a \emph{type}.
\nocite{YeganefardB11}
We have six types:  \emph{Vehicle Function (VF)}, 
\emph{Sub Function (SF)},  \emph{End Condition (EC)},  \emph{Function
Contribution (FC)}, \emph{Trigger (TR)} and \emph{Pre Condition (PC)}.
Formally, a structured requirements specification is a tuple $(\Reqs,\parent,\type)$ consisting of a set
of requirements $\Reqs$, 
a partial function
$\parent$ which maps a requirement to its parent requirement 
if there is one, and a function
$\type:\Reqs\rightarrow\Types$ which assigns  to each requirement a
type in $\Types$, where 
$\Types = \{ \mathit{VF}, \mathit{SF}, \mathit{EC}, \mathit{FC}, \mathit{TR},
\mathit{PC}\}$.

We will next explain how one derives the logical dependencies between requirements from the
hierarchical relationship $\parent$ and the classification through
types $\type$.
Intuitively, the logical dependencies between requirements
reflect the
hierarchical dependency and
the temporal dependency (the \emph{function sequence})
according to the function expressed by the type of the requirement
 (see Figure~1).
\begin{figure} [b]
\centering
\includegraphics[width=0.45\textwidth]{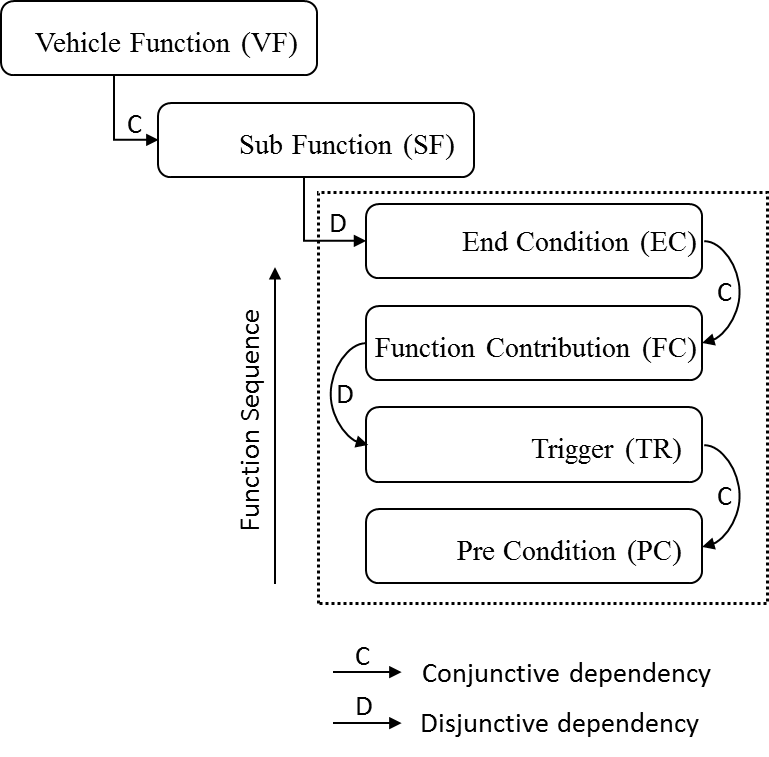}
\caption{Conjunctive vs.\ disjunctive logical
  dependencies between requirements
  according to their type.}
\label{fig:Function_Sequence}
\end{figure}

\bigskip

\noindent
\textbf{Requirements Specification $\cal R$.} \\
We introduce a Boolean variable  $\req$  for each requirement in $\Reqs$.  We will express the logical dependencies between
requirements through a logical formula $\cal R$.  Formally,   $\cal R$ is the conjunction
of all implications  of the form $\req_0{\implies}\req_1$ where:
\begin{itemize}
\item[]  $\req_0=\parent(\req_1)$,
\item[] $\type(\req_0)=\textit{VF}$, and
\item[] $\type(\req_1)=\mathit{SF}$
 \item[or]
\item[] $\parent(\req_0)=\parent(\req_1)$,
\item[] $\type(\req_0)=\textit{TR}$, and
\item[] $\type(\req_1)=\textit{PC}$
\item[or]
\item[] $\parent(\req_0)=\parent(\req_1)$,
\item[] $\type(\req_0)=\textit{EC}$, and
\item[] $\type(\req_1)=\textit{FC}$.
\end{itemize}
and all implications of the form
$\req_1{\implies}(\req_1\vee\ldots\vee\req_n)$ 
where
\begin{itemize}
\item[] $\parent(\req_0)=\parent(\req_1)=\ldots=\parent(\req_n)$,
\item[] $\type(\req_0)=\textit{FC}$, and
\item[]    $\type(\req_1)=\ldots=\type(\req_n)=\textit{TR}$
\item[or]
\item[] $\parent(\req_0)=\parent(\req_1)=\ldots=\parent(\req_n)$,
  \item[] $\type(\req_0)=\textit{SF}$, and
  \item[]    $\type(\req_1)=\ldots=\type(\req_n)=\textit{EC}$.
\end{itemize}
\bigskip

\noindent
Thus, the formula $\cal R$
is the
conjunction of  implications of the form:
\[
\req_0 {\implies} (\req_1\vee\ldots\vee\req_n)
\]
where $n\geq1$.  
We call an implication of the form above a \emph{conjunctive} logical
dependency if $n = 1$ and a \emph{disjunctive}
logical dependency if $n>1$.   The terminology `conjunctive' stems
from the fact that we express an implication of the form
$\req_0{\implies}\req_1\wedge\ldots\wedge\req_m$
by $\req_0{\implies}\req_1$, \ldots, and  $\req_0{\implies}\req_m$.

\subsection{Identifying Redundant Test Cases}
\label{subsec:test_suite_reduction}

The development process consists of a sequence of releases. Each release
represents a specific development state and has its own testing phase with a
given set of requirements and a corresponding test suite. We assume that test
cases and requirements are linked with each other~\cite{BauerK11}. Each single
requirement in the set of observed requirements for a specific release should be
covered by a test case from the corresponding test suite at least once.
Formally, we use a function that maps a test case to a set of requirements.
\bigskip

\noindent \textbf{Test Suite $\TestSuite$.} \\
A test suite $(\TCs,\link)$ for a given set of requirements
$\Reqs$ consists of a set of test cases $\TCs$ and a function
\[\link:\TCs\rightarrow2^\Reqs\] which maps a test case to a set of
requirements. Given a test suite $(\TCs,\link)$ we use the test cases $\tc \in
\TCs$ as Boolean variables and we define the \emph{test suite}
$\TestSuite$ as a logical formula in the set of Boolean variables $\TCs$, namely
as a conjunction of equivalences of the form
\[
\tc {\,\Leftrightarrow\,}
\req_1\wedge\ldots\wedge\req_n
\]
for every $\tc \in \TCs$ such that 
\[
\link(\tc)=\req_1,\ldots,\req_n
\]
holds.
\bigskip

\noindent \textbf{Test Platform  $\Platform$.}  \\
Test platforms are ordered according to their level. Intuitively, if a
requirement is satisfied at a given level, then it is also satisfied at a lower
level, but not necessarily on a higher level (the number of potential,
non-modeled error causes rises with each higher level). Consequently, if a
requirement is not satisfied at a given level, then it is also not satisfied at
a higher level, but not necessarily on a lower level.

We define the \emph{test platforms} $\Platform$ as a logical formula in the set
of Boolean variables $\Reqs$, namely as a conjunction of implications \[ \req_0
{\implies}\req_1.
\] We have the implication of the form above whenever the two requirements
$\req_0$ and $\req_1$ express the same condition but refer to different
test platforms, i.e., $\req_0$ refers to a higher test platform than $\req_1$.
\bigskip

\noindent \textbf{Test Status $\cal S$.} \\
A test status
$(\Success,\Fail,\NoResult)$ for a given set of test cases $\TCs$ consists of a triple of subsets (the subsets
contain the test cases that have \emph{succeeded}, \emph{failed}, or
that were not yet executed, thus have \emph{no result}). Given a test status
$(\Success,\Fail,\NoResult)$ we use the test cases $\tc \in \TCs$ as Boolean
variables and we define the \emph{test status} $\TestStatus$ as a logical
formula in the set of Boolean variables $\TCs$, namely as equalities that bind
the Boolean value of those test cases that have succeeded respectively failed so
far.  Formally,

\[
\begin{array}{ll}
\tc \Leftrightarrow \mathit{true} & \mbox{if \ } \tc\in\Success 
\\
\tc\Leftrightarrow \mathit{false} & \mbox{if \ } \tc\in\Fail.
\end{array}
\]
From now on, when we use a
test case $\tc$ (i.e., an element of $\TCs$) in a logical formula, it
denotes the corresponding Boolean variable.
\bigskip

\noindent \textbf{Redundancy of a Test Case.}\\ 
Given the structured requirements specification $\ReqSpec$, the
test suite $\TestSuite$, the test platforms $\Platform$, and given the current
test status $\TestStatus$, a test case $\tc$ is redundant if its result can be
inferred from $\ReqSpec$, $\TestSuite$, $\Platform$ and $\TestStatus$, formally
if either
\[
\begin{array}{ll}
&\ReqSpec \wedge \TestSuite \wedge \Platform \wedge \TestStatus\models \tc=
\mathit{true}
\bigskip
\\
\mbox{or }&\ReqSpec \wedge \TestSuite \wedge \Platform \wedge \TestStatus\models \tc=
\mathit{false}\mbox{.}
\end{array}
\]
The condition means the value of (the Boolean variable corresponding to) the test
case $\tc$ is fixed  in every model of
$\ReqSpec \wedge \TestSuite \wedge \Platform \wedge \TestStatus$.  In other words,  the value of
$\tc$ must always be $\mathit{true}$ or it must always be $\mathit{false}$ in every valuation of (the
Boolean variables corresponding to) the test cases in $\TCs$ and the
requirements in $\Reqs$ which satisfies
\begin{itemize}
\item the  logical dependencies derived from the structured
  requirements specification $\ReqSpec$,
\item the equivalences defined through the test suite $\TestSuite$,
\item the  logical dependencies defined through the test platforms
$\Platform$, and
\item the equalities defined through the current status $\TestStatus$,
  i.e., the equalities binding the Boolean values of those test
  cases that have succeeded respectively failed so far.
\end{itemize}

Since the validity of entailment can be reduced to non-satisfiability, the
condition amounts to the fact that either $\tc= \mathit{true}$  or $\tc=
\mathit{false}$ is unsatisfiable in conjunction with the implications from
$\ReqSpec$, the equivalences from $\TestSuite$, the implications from
$\Platform$, and the equalities from $\TestStatus$. Thus, the redundancy of each
test case can be inferred with an off-the-shelf SAT solver.
The practical potential of the approach for reducing the number of
test cases has been demonstrated in~\cite{ArltMo15}.

\section{Generating Test Plans}
\label{sec:GTP}

In the previous section we have described how to identify the redundancy of a
test case based on the present status of test results. In this section we give
an algorithm to generate an optimal test plan (a sequence of test cases to be
executed). The input of the algorithm is a) the dependencies between test
results (dependencies entailed from $\ReqSpec,\TestSuite$ and $\Platform$ e.g.
``$\tc_0$ implies $\tc_1$'') and b) the user's expectation on the outcome of the
individual test cases (``I expect that $\tc_0$ succeeds''). Optimality here
refers to the number of ordering constraints (``$\tc_0$ comes before $\tc_1$'') that are
entailed from a) and b).

\begin{figure}[t] \centering
\includegraphics[width=0.40\textwidth]{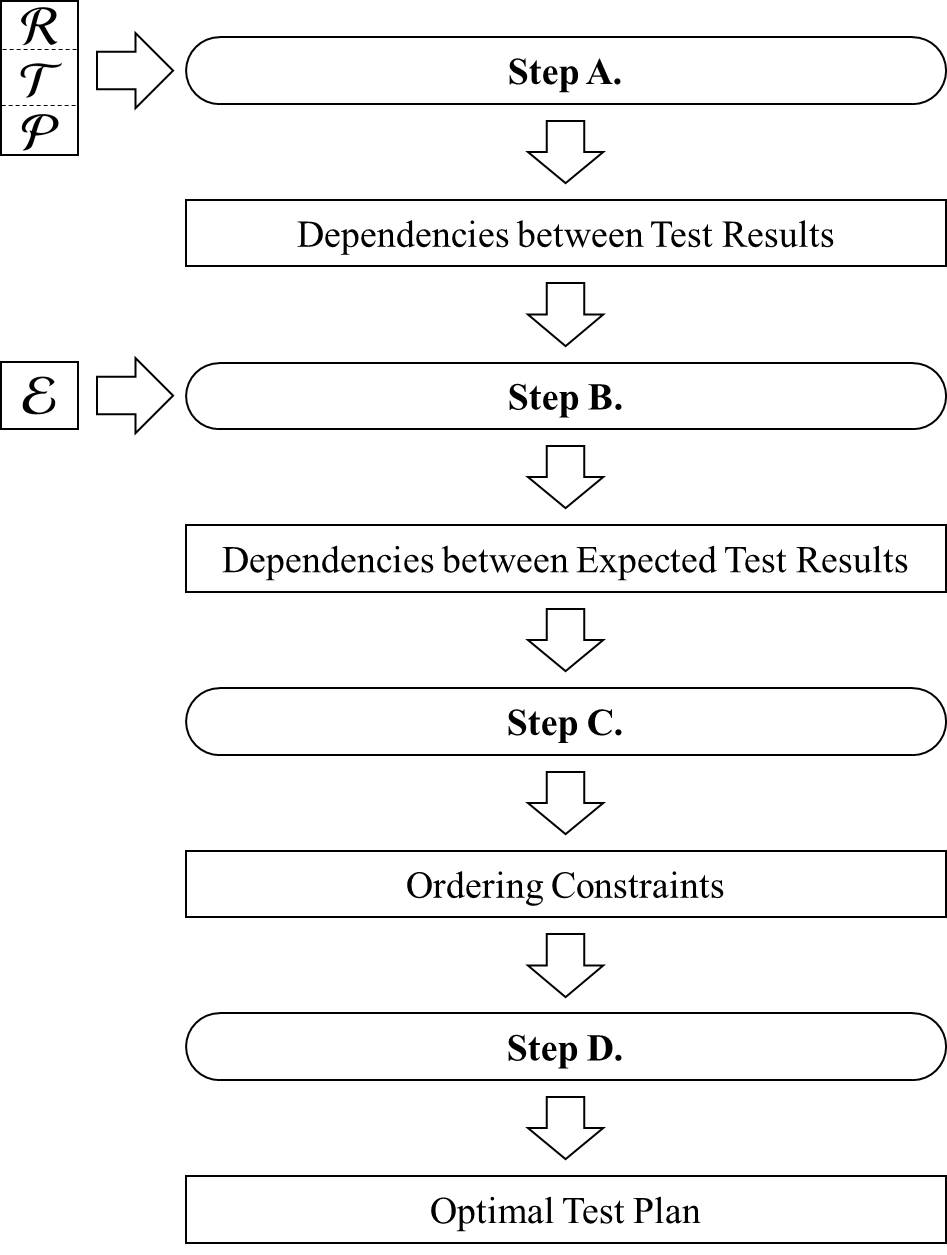}
\caption{Algorithm for generating test plans. \ The inputs are the requirements
specification $\ReqSpec$, the test suite $\TestSuite$, the test platforms
$\Platform$, and the expectation on test results $\As$.}
\label{fig:TPG_Steps}
\end{figure}

The algorithm consists of the following four steps (see
Figure~\ref{fig:TPG_Steps}). The numbering A.
- D. corresponds to the following subsection that explains the steps.

\begin{itemize}
\item [\textit{A.}] Compute the set of \emph{dependencies
between test results} that are entailed by the requirements specification $\ReqSpec$, the
test suite $\TestSuite$ and the test platforms $\Platform$. For example, this step could return the logical implication
$\tc_0 \implies \tc_1$.
\item [\textit{B.}] 
Compute the set of dependencies between the expected test
results that are entailed by the requirements specification $\ReqSpec$, the
test suite $\TestSuite$, the test platforms $\Platform$, the test status
$\TestStatus$ and the \emph{user's expectation} $\As$. For example, this step
could return the logical formula $\neg\ell_0 \implies \neg\ell_1$ where $\neg\ell_0$
stands for the fact that the user expects a negative test results for $\tc_0$.
\item [\textit{C.}] First, infer which test result becomes redundant from what
set of expected test results. For example, this step could return that the test
result of $\tc_1$ is redundant from the expected (negative) test result of $\tc_0$.
Then, derive \emph{ordering constraints} between test cases. For
example this step could return that $\tc_0$ should be scheduled before $\tc_1$.
\item [\textit{D.}] \emph{Generate a test plan}. For example, the test plan
could schedule $\tc_0$ before $\tc_1$.
\end{itemize}

For simplicity we took a simple example of a logical implication of the form
$\tc_0 \implies \tc_1$. As already indicated in the introduction, step A. can
return more complex formulas than just a logical implication as in the example.
\bigskip

\subsection{Dependencies between Test Results}
\label{sec:Dep_betw_TR}

\noindent \textbf{Informal Discussion.}\\
In this section we will explain how one can compute the dependencies between
test results as discussed in the example in the introduction, in the scenario
of the rain sensor. Formally, the
dependencies are modelled by logical formulas that are entailed by the logical
formulas that model the requirements, the test links, and the test platforms
(see Section~\ref{sec:preliminaries}).
\bigskip

\noindent \textbf{Formal Description.} \\
We now give the algorithm to compute
all possible dependencies between test results. We will start by formally defining
the notion of dependency between test results.
\bigskip

\noindent \textbf{Definition 1} [Dependencies between Test Results] \\
We call a formula of the form
\[
\tc_1 \wedge\ldots\wedge \tc_n \implies \tc_{n+1} \vee\ldots\vee \tc_m
\]
a \emph{dependency between test results} (here, between the results of the test
cases $\tc_1,\ldots,\tc_n,\tc_{n+1},\ldots,\tc_m$).
\bigskip

The following remark expresses the relevance of the notion defined
in Definition 1.
\bigskip

\noindent \textbf{Remark 1} [Dependencies between Test Results] \\
If a dependency between the $n+m$ test cases
$\tc_1,\ldots,\tc_n,\tc_{n+1},\ldots,\tc_m$ is entailed by the requirements specification $\ReqSpec$, the
test suite $\TestSuite$, and the test platforms $\Platform$, then 
each of the $n+m$ test cases can be made redundant by the other ones.
\bigskip

By ``$\tc$ can be made redundant'' we mean that there exists a test status
$\TestStatus$ such that $\ReqSpec, \TestSuite, \Platform$ and $\TestStatus$
entail the result of $\tc$, and thus $\tc$ is redundant if all the other test cases ($n+m-1$ many) are scheduled before.
If the other test cases have been executed, then the test status $\TestStatus$
assigns a Boolean value modelling their test results.

Often, each disjunction or its negation is called a \emph{clause} and each
disjunction (a Boolean variable $\tc$ or its negation $\neg\tc$) is called a
\emph{literal}. The conjunction is often written as a \emph{set}; i.e., the
resulting formula is a set of clauses.
\bigskip

\noindent{\bf{Algorithm to Compute Dependencies between Test Results.}} \\
We next explain how we compute the set of all dependencies that are entailed by
$\ReqSpec,\TestSuite$ and $\Platform$.

As explained in Section~\ref{sec:preliminaries}, we use the requirements
specification $\ReqSpec$ as a conjunction of implications of the form
\[
\req {\implies} (\req_1\vee\ldots\vee\req_n)
\]
for every Boolean variable $\req \in \Reqs$.
The test suite $\TestSuite$ is a conjunction of equivalences of the form
\[
\tc {\,\Leftrightarrow\,} \req_1\wedge\ldots\wedge\req_n
\]
for every Boolean variable $\tc \in \TCs$.
The test platforms $\Platform$ are implications of the form
\[
\req_0 \implies \req_1.
\]
Thus the conjunction
\[
\ReqSpec\wedge\TestSuite\wedge\Platform
\]
is a Boolean formula in variables from the set $\Reqs \cup \TCs$, i.e. in
variables that stand for requirements and in variables that stand
for test results.

The formula \[ \exists \Reqs (\ReqSpec\wedge\TestSuite\wedge\Platform) \] stands
for the formula with an existential quantification $\exists \req$ for every
variable $\req \in \Reqs$. 

In the first step of the algorithm, we apply a quantifier elimination
procedure and obtain an equivalent formula (without existential quantifiers) in variables only
from the set $\TCs$, i.e. in variables that stand for test results. 

In the second
step for the algorithm, we transform the resulting formula into conjunctive
normal form, i.e., into a conjunction of disjunctions, where each disjunct is of the form $\tc$ or
of the form $\neg\tc$, for $\tc \in \TCs$.

In the third and last step of the algorithm, we \emph{saturate} the set of clauses; i.e., we add clauses that
are derivable under the \emph{resolution rule} until no more new clauses can be
added. For example, given the clauses $\tc_0\vee \tc_1$ and
$\tc_0\vee \neg\tc_1$, we will derive and add the clause $\tc_0$.

The resulting set is then exactly the set of dependencies
(dependencies between test
results)  that are entailed by
$\ReqSpec,\TestSuite$ and $\Platform$.  That is, 
a dependency between test results is entailed by
$\ReqSpec,\TestSuite$ and $\Platform$ if and only if
it lies in the set.
\bigskip

\subsection{Dependencies between Expected Test Results}
\label{sec:UE}

\noindent \textbf{Informal Discussion.}\\
We come back to the example discussed in the introduction.  The
example indicates that already a test plan with an ordering of only
two test cases cannot be optimal independently of the results of the
test cases.  We have the test case $\tc_0$ for the requirement \emph{sunroof
closes automatically when it is raining} and the test cases $\tc_1$
for the requirement \emph{rain sensor detects rain}, and we have 
the dependency between the two test results $\tc_0 \vee \neg\tc_1$, formally
\[
\ReqSpec \wedge \TestSuite \wedge \Platform \models \tc_0 \vee \neg\tc_1.
\]

If the user expects that both test cases are going to succeed (both boolean
variables are \emph{true}) then the test plan should start with the execution of $\tc_1$.

If the user expects that both test cases are going to fail (both boolean
variables are \emph{false}) then the test plan should start with the execution of $\tc_0$.

We now come back to Remark 1. The remark says that each test case $\tc$
that appears among the $n+m$ test cases involved in a dependency of
the form 
\[
\tc_1 \wedge\ldots\wedge \tc_n \implies \tc_{n+1} \vee\ldots\vee \tc_m 
\] 
can be
made redundant if all the other test cases ($n+m-1$ many) are scheduled before
and their results are corresponding, i.e., if the test status $\TestStatus$
models their results, then $\ReqSpec, \TestSuite,\Platform$ and
$\TestStatus$ entail the result of $\tc$.

Obviously, the user is not able to make a prophecy of the test results, i.e.,
she does not know $\TestStatus$ before executing the test cases. However, the
user has an expectation on the test results and she wants to know what test case
$\tc$ will be redundant if the other test cases are scheduled before and the
test results are according to her expectation.
\smallskip

We next introduce the concept of a \emph{default specification}.  As
mentioned in the introduction, the test manager can
use a default specification to specify her
expectation (and can  take it as is
or take it as a basis, changing individual test cases 
according to her personal insight and experience).  
\\
\emph{(1) Each test case will fail.}  This pessimistic default
expectation seems applicable at an early level of maturity, towards
the beginning of a development process.\footnote{It is interesting
  here to
  contrast the initial expectation in software testing (the software
  will be buggy) with, say, the
  initial expectation in a medical check-up (the patient will be healthy).} 
\\
\emph{(2) Each test case will succeed.}  This
optimistic default expectation seems applicable at an advanced level
of maturity, towards the end of a development process.
\\
\emph{(3) Each test case will have the same result as the previous
  time.}  This is the default expectation that seems applicable at an
intermediate level of maturity.

As
mentioned in the introduction, the test manager may also leave unspecified her
expectation on a test case.
An unspecified expectation (or one that turns out to be wrong) may
possibly lead to missing a redundancy (in the case where the redundancy could have
been inferred otherwise) but our
formal criterion for the optimality of a test plan will still apply.
\smallskip

\noindent \textbf{Formal Description.}\\
We assume that the user has formalized her expectation on the test results as a
function $\Expf$ that maps each test case $\tc$ to \textit{success} or
to \textit{fail}. For each test case $\tc \in \TCs$ we introduce a Boolean
variable $\ell_{\tc}$ which we will use in the encoding of the user expectation
as a logical formula. Notation: Given $\tc_i$, we will write $\ell_i$ instead of
$\ell_{\tc_i}$.
\bigskip

\noindent \textbf{Definition 2} [Expectation on Test Results $\As$]\\
The expectation on test results is a logical formula $\As$ over the set of
Boolean variables $\{\ell_{\tc}\mid \tc \in \TCs\}$, namely a conjunction of
equivalences. The equivalence $\ell_i \leftrightarrow \tc_i$ encodes that user
expects a \emph{positive} test result for $\tc_i$. The equivalence $\ell_i
\leftrightarrow \neg \tc_i$ encodes that user expects a \emph{negative} test
result for $\tc_i$.
\bigskip

Using the convention that a set denotes the conjunction of
its elements, we can write:
\[
\begin{array}{r@{}r@{}l@{}}
  \As = &\{\ell_{\tc} \leftrightarrow \tc &\mid \tc \in \TCs, \Expf(\tc) =
  success\}\\
  \cup &\{\ell_{\tc} \leftrightarrow \neg \tc &\mid \tc \in \TCs, \Expf(\tc) =
  fail\}.
\end{array}
\]
\smallskip

\noindent \textbf{Remark 2} [Expectation on Test Results $\As$]\\
The variable $\ell_\tc$ has the value $true$ if and only if the value of $\tc$
corresponds to the expected result. Formally, if \emph{b} stands for the Boolean
constant $true$ if $\Expf(\tc)=success$, and for the Boolean constant $false$
if $\Expf(\tc)=fail$, then we have:
\[
\As \models \ell_\tc \Leftrightarrow (\tc \Leftrightarrow \emph{b})
\]

Introducing the concept of the \emph{expectation on test results} is the key
idea in our paper.
Adding the formula $\As$ to the formulas $\ReqSpec$, $\TestSuite$, and
$\Platform$ will allows us to compute what test case $\tc$ can be inferred as
redundant (and potentially be dropped) and what test cases have to be scheduled
before $\tc$ (because the result of $\tc$ can be inferred from the results of
those test cases), always under the assumption that the results of those test
cases are as expected.
\bigskip

\noindent{\bf{Algorithm to Compute Dependencies between Expected Test
Results.}}\\
We can now give the algorithm to compute all dependencies between
expected test results that are entailed by $\ReqSpec,\TestSuite, \Platform$, and $\As$. We
first apply the algorithm described in Section~\ref{sec:Dep_betw_TR} and compute the set of all dependencies between
test results. We transform each dependency between test results (an implication)
into a disjunction, i.e., a clause.
For each (positive or negative) occurence of $\tc$ in a clause we replace $\tc$
by $\ell_{\tc}$ if $\Expf(\tc)$ is equal to $success$ and by $\neg
\ell_{\tc}$ if $\Expf(\tc)$ is equal to $fail$. We then obtain a set of clauses
over variables $\ell_{\tc}$ where $\tc \in \TCs$. We now transform each clause
into an equivalent implication.

The resulting set is then exactly the set of dependencies
(dependencies between expected test
results)  that are entailed by
$\ReqSpec$, $\TestSuite$, $\Platform$, and $\As$.  That is,
a dependency between expected test results
is entailed by
$\ReqSpec$, $\TestSuite$, $\Platform$, and $\As$, if and only if
it lies in the set.
\bigskip

\subsection{Ordering Constraints}
\label{sec:HC}

\noindent \textbf{Informal Discussion.}\\
Using the formula for computing dependencies between test results and the user's
expectation on test results we are able to automatically identify implied
test results.

An example, assume that $\ReqSpec, \TestSuite$ and $\Platform$ entails the
dependency between test results \[ \tc_2 \wedge \tc_3 \implies \tc_0 \vee \tc_1
\] which is equivalent to the clause \[ \tc_0 \vee \tc_1 \vee \neg\tc_2 \vee
\neg\tc_3.
\] Assume that the user's expectation is as follows.
\[
\begin{array}{ll}
&\Expf(\tc_0) = fail\\
&\Expf(\tc_1) = fail\\
&\Expf(\tc_2) = success\\
&\Expf(\tc_3) = fail\\
\end{array}
\]
As described in Section~\ref{sec:UE}, we encode the user's expectation in the
logical formula $\As$. We then have that $\ReqSpec,
\TestSuite, \Platform$ and $\As$ entail the dependency between expected test
results
\[
\ell_0 \wedge \ell_1 \wedge \ell_2 \implies \ell_3
\]
which is equivalent to the clause
\[
\neg\ell_0 \vee
\neg\ell_1 \vee \neg\ell_2 \vee \ell_3.
\]
We use the dependency between expected test results in order to infer the
ordering constraints $\tc_0 < \tc_3, \tc_1 < \tc_3$ and $\tc_2 < \tc_3$. The
ordering constraints tell us that we should schedule $\tc_0, \tc_1$ and $\tc_2$
before $\tc_3$. Indeed, if the results of $\tc_0, \tc_1$ and $\tc_2$ are as
expected then $\tc_3$ is redundant (its result can be inferred from $\ReqSpec, \TestSuite$ and $\Platform$ and the result of $\tc_0, \tc_1$ and
$\tc_2$).

The above example shows that we can infer ordering constraints from a dependency
between expected test results if the dependency is expressed by an implication
which has only one disjunct on the right hand side (this is equivalent to the
fact that the corresponding clause has exactly one positive disjunct). The next
example shows that we cannot infer ordering constraints when there is
more than one disjunct in the right hand side of the implication (if the clause has
more than one positive disjunct).

Assume now that the user's expectation is as follows.
\[
\begin{array}{ll}
&\Expf(\tc_0) = success\\
&\Expf(\tc_1) = success\\
&\Expf(\tc_2) = success\\
&\Expf(\tc_3) = success\\
\end{array}
\]
We then have that $\ReqSpec,
\TestSuite, \Platform$ and $\As$ entail the dependency between expected test
results
\[
\ell_2 \wedge
\ell_3 \implies \ell_0 \vee \ell_1
\]
which is equivalent to the clause
\[
\ell_0 \vee
\ell_1 \vee \neg\ell_2 \vee \neg\ell_3.
\]
Now, the dependency between expected test results does not allow us to infer any
ordering constraints. In fact, there exists no execution order between $\tc_0,
\tc_1, \tc_2$ and $\tc_3$ where one of the four test results could be inferred
by the three other ones (if the three other ones are executed before and their
results are as expected).
\bigskip

\noindent \textbf{Formal Description.}\\
We will next define the notion of an ordering constraint and we will present the
algorithm to infer ordering constraints.
\bigskip

\noindent \textbf{Definition 3} [Ordering constraint]\\
An ordering constraint is a conjunction of inequalities of the form
\[
\tc_0 < \tc \wedge \ldots \wedge \tc_n < \tc
\]
between the test cases $\tc_0$, \ldots, $\tc_n$ on the left-hand side and the test case
$\tc$ on the right-hand side.

A test plan satisfies the ordering constraint of the form above if the test
plan schedules $\tc_0, \ldots, \tc_n$ before $\tc$ (i.e., in the
sequential ordering specified by the test plan, $\tc_0, \ldots, \tc_n$ occurs before $\tc$).
\bigskip

\noindent \textbf{Algorithm to Compute Ordering Constraints.} \\
We can now give
the algorithm to compute all ordering constraints from a given requirements
specification, a given test suite, the test platforms, and the user's
expectation, modelled by $\ReqSpec$, $\TestSuite$, $\Platform$, and $\As$,
respectively. 

In the first step of the algorithm, we apply the algorithm described in
Section~\ref{sec:Dep_betw_TR} and compute all dependencies between expected
test results. 

In the second step, we take the set of all dependencies between expected test
results that can be expressed by a \emph{Horn clause}, i.e., an implication with
exactly one disjunct on the right hand side (equivalently a clause with exactly
one positive disjunct).   We compute its subset of \emph{minimal}
dependencies by eliminating each dependency that is subsumed by
another one in the set.  A first dependency $\tc_1 \wedge\ldots\wedge \tc_n
\implies \tc$ is subsumed by a second dependency $\tc_1 \wedge\ldots\wedge \tc_m \implies
\tc$ if the set of conjuncts of the first is contained by the second,
i.e., if $\{\tc_1,\ldots, \tc_n\}\subseteq\{\tc_1,\ldots, \tc_m\}$.

In the third and final step of the algorithm, we form
the ordering constraint
\[
\tc_1 < \tc \wedge \ldots \wedge \tc_n < \tc
\]
for each minimal dependency
\[
\ell_1 \wedge\ldots\wedge\ell_n \implies
\ell_{\tc}
\]
in the subset
obtained after the second step (which eliminates each dependency that
is not minimal).
\bigskip

\noindent \textbf{Theorem} [Completeness of the Algorithm]\\
For every test case $\tc$, if
there exists an ordering of test cases in which the result of $\tc$
becomes redundant, then the algorithm will infer such an
ordering.
More precisely, the algorithm will infer an ordering constraint such that
 the test case $\tc$ becomes redundant in every sequence of test cases
 (with the expected results)
 that satisfies the ordering constraint.
\smallskip

\noindent \textbf{Proof} [Completeness of the Algorithm]\\
We assume that a test result $\tc$ is redundant in the sequential
ordering $\tc_1,\ldots,\tc_n$ followed by $\tc$.
Moreover, we assume that $\{\tc_1,\ldots,\tc_n\}$ is a \emph{minimal} set of
test cases;  i.e., if we dropped one test case from the set, then
$\tc$ would no longer be redundant.
We need to show that our algorithm
will infer the ordering constraint $
\tc_1 < \tc \wedge \ldots \wedge \tc_n < \tc
$
which are satisfied by the minimal sequential
ordering $\tc_1,\ldots,\tc_n,\tc$.

The redundancy means that the result of the test case $\tc$ (say, $false$)
can be inferred from $\ReqSpec$, $ \TestSuite$,  $ \Platform$, and
from the present test status $\TestStatus$ after executing the
sequence of test cases  $\tc_1,\ldots,\tc_n$ and right before executing the
test case $\tc$.  
That is, we have
\[
\ReqSpec \wedge \TestSuite \wedge \Platform \wedge \TestStatus \models
\tc = false.
\]
The test status $\TestStatus$ fixes the Boolean value $b_i$ for each of the
test cases $\tc_1,\ldots,\tc_n$.
That is, $\TestStatus$ is equivalent to the conjunction of the equivalences
$\tc_i\Leftrightarrow b_i$, formally
\[
\TestStatus \equiv \tc_1\Leftrightarrow b_1\wedge\ldots\wedge \tc_n\Leftrightarrow b_n.
\]
Thus, we have
\[
\ReqSpec \wedge \TestSuite \wedge \Platform \models
(\tc_1\Leftrightarrow b_1)\wedge\ldots\wedge (\tc_n\Leftrightarrow b_n)
\Rightarrow \tc = false.
\]
Since we assume that the execution of the test cases
$\tc_1,\ldots,\tc_n$ produces the expected result, and since the
equivalence $\ell_i\Leftrightarrow \tc_i$ lies in $\As$ whenever $b_i$
is equal to $true$, and the
equivalence $\ell_i\Leftrightarrow \neg\tc_i$ lies in $\As$ whenever $b_i$
is equal to $false$, we have that (see also Remark 2)
\[
\As\models\ell_i\Leftrightarrow ( \tc_i\Leftrightarrow b_i).
\]
Thus, we have that
\[
\ReqSpec \wedge \TestSuite \wedge \Platform \wedge \As \models
\ell_1\wedge\ldots\wedge \ell_n
\Rightarrow \tc = false.
\]
Since we assume that the execution of \emph{all} test cases (i.e., including the
redundant test case $\tc$) produces the result corresponding to the
expectation and the result of $\tc$ can already be inferred to be
$false$ from the results of the test cases preceding $\tc$, we also have that $\As\models \ell_\tc\Leftrightarrow
\neg\tc$ and hence $\As\models \ell_\tc\Leftrightarrow
 ( \tc\Leftrightarrow false)$.  Thus, we have that
\[ 
\ReqSpec \wedge \TestSuite \wedge \Platform \wedge \As \models
\ell_1\wedge\ldots\wedge \ell_n
\Rightarrow \ell_{\tc}.
\]
By our assumption on the minimality of the set of
test cases
$\{\tc_1,\ldots,\tc_n\}$ (i.e., if we dropped one test case from the set, then
$\tc$ would no longer be redundant), we have that the Horn clause
above is minimal (i.e., if we dropped one of the conjuncts from the
body, the implication would no longer be entailed by $\ReqSpec \wedge
\TestSuite \wedge \Platform \wedge \As$).  

As a consequence, our algorithm  infers the Horn clause above as a
dependency between expected test results in Step~B, and the algorithm
will infer the ordering constraint
\[
\tc_1 < \tc \wedge \ldots \wedge \tc_n < \tc
\]
which are satisfied by the minimal sequential
ordering $\tc_1,\ldots,\tc_n,\tc$.  This terminates the proof.
\bigskip

\subsection{Optimal Test Plan}
\label{sec:AGTP}

\noindent \textbf{Informal Discussion.}\\
We give an example that illustrates a difficulty in the generation of a test
plan that satisfies all ordering constraints. Assume that we have the following
two dependencies between test results 
\[
\begin{array}{ll}
\tc_0 \wedge \tc_1 &\implies \tc_2\\
\tc_3 &\implies \tc_0 \vee \tc_1
\end{array}
\]
and assume the user's expectation 
\[
\begin{array}{ll}
&\Expf(\tc_0) = success\\
&\Expf(\tc_1) = fail\\
&\Expf(\tc_2) = fail\\
&\Expf(\tc_3) = success.\\
\end{array}
\]
Our algorithm computes the following ordering constraints.
\[
\begin{array}{ll}
\tc_0 < \tc_1\\
\tc_1 < \tc_0\\
\tc_2 < \tc_1\\
\tc_3 < \tc_0\\
\end{array}
\]
There does not exist a test plan that satisfies all these ordering constraints
(because there is no sequential ordering that satsifies the first two ordering
constraints, i.e., $\tc_0 < \tc_1$ and  $\tc_1 < \tc_0$).
\bigskip

\noindent \textbf{Formal Description.}\\
Since in general it is not possible to generate a test plan that satisfies all
ordering constraints computed by the algorithm, we can only ask for a test plan
which maximizes the number of ordering constraints that are satisfied by the
test plan.
\bigskip

\noindent \textbf{Algorithm to Compute an Optimal Test Plan.}\\
In the first step, given the set of ordering constraints computed by the
algorithm of Section~\ref{sec:HC}, we compute a subset with the maximal number
of ordering constraints that can be satisfied simultaneously.
The problem of computing the maximal number of 
sets
of ordering constraints that can be satisfied simultaneously, can be reduced to
a variant of a well known NP-complete problem, namely the problem to compute a
minimum feedback arc set in a directed graph.
As usual, each inequality in an ordering constraint is translated into an edge
of the directed graph, but now we have additional \emph{hyperedges} which go
from a set of source nodes to a (single) target node.
Namely, each ordering constraint of the form
\[
\tc_1 < \tc \wedge \ldots \wedge
\tc_n < \tc
\]
is translated to a hyperedge that goes from the set of nodes
$\{\tc_1,\ldots,\tc_n\}$ to the node $\tc$.

For efficient implementations of algorithms to solve this
problem, see e.g.,~\cite{Baharev15}.

In the second step, we compute an optimal test plan, i.e., a
sequential ordering of test cases that satisfies all ordering constraints in the
set that we have computed in the first step, a set with the maximal
number of ordering constraints that can be satisfied simultaneously.
Here, we can apply classical algorithms for topological sorting (see,
e.g.,~\cite{Ajwani12}).
\bigskip

\noindent{\bf{Online Update of a Test Plan.}}\\
If during the execution of a test plan the execution of a test case leads to a
result that is different from the expectation on the result then the user may
consider to adapt the test plan. In this case, all she has to do is to update
the function $\Expf$ (which leads to an update of the logical formula
for the expectation $\As$) and re-execute
the algorithm to generate new ordering constraints and a new optimal test plan.
\bigskip

\section{Related Work}
\label{sec:related}

Generating an optimal test plan seems related to the
prioritization of test cases, which is a topic of very active research
which we will discuss in greater detail below.

The goal of prioritization is generally phrased as obtaining a maximum amount of
information about the maturity of the system under development as early as
possible.
During the execution of the test cases, at the moment when resources have run
out, the test manager will have to drop test cases; i.e., the decision depends
on external factors.

We next discuss work on prioritization in the setting of regression testing.
The formal setup makes our approach \emph{a
  priori} independent of a particular practical setting.
However, the setup of regression testing may seem particularly suitable for our
approach because it can facilitate the task of the user to revise her
expectation of test results.

The work in~\cite{Engstrom11,Fazlalizadeh09,Kim02,Qu07,Sherriff07} uses the
history of test results for the priorization.  In contrast, the work of
\cite{Wong95,Harrold99} uses the explicit knowledge about what fault can be
revealed by what test case.  Alternatively, the priorization can be based on
coverage criteria; see, e.g.,
\cite{Elbaum00,Li07,Rothermel99,Rothermel01,Yang09}.
Another line of research investigate priorization under the header of increasing
the failure detection potential (FDP) by concentrating on the mutations the
program~\cite{Budd80,Saha15,Srivastava02}.  The work
in~\cite{Fraser07,Korel07,Korel08,Korel05} goes into the same direction, using
system models.

The work in~\cite{Srikanth05} is related to ours in that it also starts with the
requirements.  In fact, it first prioritizes the requirements and then derives
the prioritization of the test cases from the prioritization of the
requirements.

The work in \cite{Elbaum01,Walcott06} bases the prioritization on the cost of
the execution of the test cases.

The work in~\cite{Arlt12} uses dataflow analysis in order to eliminate redundant
test cases.   The dataflow analysis is done offline; i.e., the results of the
test cases are not taken into account.
\bigskip

\section{Conclusion}
\label{sec:conclusion}

As the discussion of related work shows, the work presented in this paper is quite
different from existing work, by its topic, and also in its style.

We present the formal foundation for what could become a new line of work,
namely the automatic generation of test plans from a specification (here, the
specification of the expectation on test results) and according to a criterion
for optimality.

We introduce  a novel algorithmic problem.
We give an algorithm approach to solve the problem.
The logical setting allows us to formulate the approach in a concise manner.

We formally characterize the contribution of the approach.
This is quite different from work that experimentally validates the contribution
of a new approach.

As for future work, our approach cannot be viewed stand-alone.
The automatic generation of an optimal test plan can only be the first step.
The conception of a test plan will take into account many different objectives,
the reduction of the set of test cases being only one of these.  The conception
of a test plan  will also take into account constraints on the execution of the
test plan, the cost of preparing a test case (so that, e.g., we may have sets of
test cases that must be grouped together), the availability of a testing
resource, etc.
Furthermore, the redundancy of a test case seems orthogonal to the criteria used
in the methods for prioritization discussed above.  The automatic generation of
test cases can in principle be integrated with each of these methods.
All this indicates interesting directions for future research.
\bigskip

\mbox{}

\bibliographystyle{abbrv}
\bibliography{main}

\end{document}